\documentclass[
 aps,prmaterials,
 superscriptaddress,
 amsmath,amssymb,
 reprint,%
]{revtex4-2}

\usepackage{hyperref}
\hypersetup{
    colorlinks=true,
    linkcolor=blue,
    filecolor=blue,      
    urlcolor=blue,
    citecolor=blue,
}
\usepackage{lipsum}
\usepackage{graphicx}
\usepackage{dcolumn}
\usepackage{bm}
\usepackage{upgreek}

\usepackage[utf8]{inputenc}
\usepackage[T1]{fontenc}
\usepackage{mathptmx}
\usepackage{etoolbox}
\newcommand{\BS}{Bi$_2$Se$_3$}
\newcommand{\keff}{$K_{\text{eff}}$}

\begin{document}

\AtEndEnvironment{thebibliography}{
\bibitem{data}
        B. A. Brereton, S. Hait, A. Yagmur, C. J. Kinane, F. Maccherozzi, M. Conroy,
        S. Sasaki, T. A. Moore, S. S. Dhesi, S. Langridge and C. H. Marrows, Research Data Leeds, https://doi.org/TBD, 2025

\bibitem{Supp}
  See Supplemental Material at [URL will be inserted by publisher] for for details.

}

\title{
Tailoring Ultrathin Magnetic Multilayers at Terraced Topologically Insulating Interfaces\\ for Perpendicularly Magnetized Domains}

\author{Benjamin~A.~Brereton}
\affiliation{School of Physics and Astronomy, University of Leeds, Leeds LS2 9JT, UK}
\affiliation{ISIS Neutron and Muon Source, STFC Rutherford Appleton Laboratory, Harwell Science and Innovation Campus, Chilton, Oxfordshire OX11 0QX, UK}
\affiliation{Diamond Light Source, Harwell Science and Innovation Campus, Chilton, Oxfordshire OX11 0DE, UK}

\author{Soumyarup~Hait}
\affiliation{School of Physics and Astronomy, University of Leeds, Leeds LS2 9JT, UK}

\author{Ahmet~Yagmur}
\affiliation{School of Physics and Astronomy, University of Leeds, Leeds LS2 9JT, UK}

\author{Christy~J.~Kinane}
\affiliation{ISIS Neutron and Muon Source, STFC Rutherford Appleton Laboratory, Harwell Science and Innovation Campus, Chilton, Oxfordshire OX11 0QX, UK}

\author{Francesco~Maccherozzi}
\affiliation{Diamond Light Source, Harwell Science and Innovation Campus, Chilton, Oxfordshire OX11 0DE, UK}

\author{Michele~Conroy}
\affiliation{Department of Materials, Imperial College London, Exhibition Road,
London SW7 2AZ, UK}

\author{Satoshi~Sasaki}
\affiliation{School of Physics and Astronomy, University of Leeds, Leeds LS2 9JT, UK}

\author{Thomas~A.~Moore}
\affiliation{School of Physics and Astronomy, University of Leeds, Leeds LS2 9JT, UK}

\author{Sarnjeet~S.~Dhesi}
\affiliation{Diamond Light Source, Harwell Science and Innovation Campus, Chilton, Oxfordshire OX11 0DE, UK}

\author{Sean~Langridge}
\affiliation{ISIS Neutron and Muon Source, STFC Rutherford Appleton Laboratory, Harwell Science and Innovation Campus, Chilton, Oxfordshire OX11 0QX, UK}

\author{Christopher~H.~Marrows}
\email{c.h.marrows@leeds.ac.uk}
\affiliation{School of Physics and Astronomy, University of Leeds, Leeds LS2 9JT, UK}

\date{\today}

\keywords{Topological Insulators, Skyrmions, Magnetic Anisotropy, Polarized Neutron Reflectometry, PEEM}

\begin{abstract}
Topological insulators and skyrmion-hosting, chiral magnetic multilayers are two well-explored areas of modern condensed matter physics, each offering unique advantages for spintronics applications. In this paper, we demonstrate the optimization process for the growth of a \BS/buffer/[Pt/CoB/Ru]$_{\times N}$ heterostructure that combines these two material classes: the \BS\ epilayer was grown by molecular beam epitaxy before transfer under ultrahigh vacuum to a separate growth chamber where the polycrystalline metallic multilayer was sputter deposited. The structure of the samples was characterized by co-fitted X-ray and polarized neutron reflectometry measurements and scanning transmission electron microscopy. Polarized neutron models and standard magnetometry show that a buffer layer exceeding a critical thickness is required to obtain the desired uniform, perpendicular magnetic anisotropy in every magnetic layer in the multilayer. Samples with both Ta and Mo buffers were used requiring thicknesses of 1.5 and 0.9~nm respectively. In minimizing the \BS\ terracing, buffered samples yield well-defined, out-of-plane, magnetic domains suitable for spin-orbit torque induced manipulation as determined by X-ray photoemission electron microscopy. 
\end{abstract}
                          
\maketitle

\section{Introduction}

Topological insulators (TIs) have long been of interest in next-generation spintronic devices because of their highly efficient charge-to-spin conversion, analogous to the spin Hall angles of heavy metals: $\theta_{\text{SH}} = J_\mathrm{s}/J_\mathrm{c}$. In such systems $\theta_{\text{SH}}$ can be much greater than unity \cite{Mellnik2014,Fan2016}, arising from spin-momentum locked surface states. Such devices utilize the large spin-orbit torques (SOTs) generated by this effect, and often are based on switching thin ferromagnetic films with perpendicular magnetic anisotropy (PMA) for low-current magnetic random access memory (MRAM) applications \cite{Zhang2025,Fan2014,Wang2017,Han2017}. 

Magnetic multilayers (MMLs) comprising ultrathin layers of magnetic and heavy metals, are known to host multitudes of exotic, chiral spin textures in systems of broken inversion symmetry due to a non-vanishing interfacial Dzyaloshinskii–Moriya interaction (DMI): most notably N\'{e}el-type magnetic skyrmions \cite{Heinze2011,Moreau,Woo2016,Chen2015,Soumyanarayanan2017,Jiang2015}. Skyrmions are topologically non-trivial with a resultant quantized Berry phase for transport electrons: their topology granting them relative stability against perturbations and admitting SOT-driven motion and manipulation from spin currents such as those generated by the spin-Hall effect (SHE) in heavy metals \cite{Hamamoto2015,Woo2016,Romming2013,Jiang2017,Zeissler2020, Buttner2017}. These properties have not gone overlooked for potential applications in energy efficient data storage and computation applications: most recently in reservoir computing \cite{Fert2013,Fert2017,Pinna2020,Yokouchi2022}. TI/chiral magnetic multilayer heterostructures present an opportunity to utilise the strengths of both materials to further reduce necessary current densities and enhance energy efficiency.

Often chiral multilayers rely on the interfacial hybridization between Co and Pt. This yields both strong PMA and DMI effects. The total effective anisotropy, $K_{\text{eff}}$, can be expressed as a function of ferromagnet (FM) thickness, $t_{\text{FM}}$, as

\begin{equation}
    K_{\text{eff}} =  \frac{K_{\text i}}{t_{\text{FM}}} + \frac{1}{2}\mu_0 M_{\text s} H_{\text D} \, , \label{eq:keff}
\end{equation}

where $K_{\text i}$ is the interfacial PMA term, $M_\mathrm{s}$ the saturation magnetization, and $H_\mathrm{D}$ the demagnetization field. In ultrathin films the demagnetization field arising from shape anisotropy, $H_\text{D} = - M_\text{s}$. The quality of the Pt/Co interface plays a key role in determining the magnitudes of the DMI vector and $K_\text{i}$ \cite{Bersweiler2016,Wells2017,Hrabec2014}, which increasingly dominates over the other volumetric contributions at low $t_{\text{FM}}$. The DMI contribution competes against a standard Heisenberg-like exchange interaction producing non-collinear and incommensurate N\'{e}el-type spin-canting over a characteristic lengthscale, $\Delta$.  

Previous PMA TI/FM switching studies \cite{Pan2022,Dc2018} note their use of buffer layers deposited on the TI surface before the MML is grown for out-of-plane (OOP) switching geometries and, when buffer layers are removed, FM easy axes tend to lie in-plane (IP). Inserting a buffer layer over a TI will introduce spin-flip scattering causing the spin-signal to be lost over a material dependent spin-diffusion length $l_{\text{SD}}$ according to $J_s(t) = J_{s,0}\,e^{-t/l_{SD}}$; this is in addition to the additional scattering introduced as a result of introducing a second interface \cite{Sanchez2014}. This problem is particularly acute for heavy, refractory metals such as Ta, which possess large spin-orbit coupling (SOC) and hence spin diffusion lengths are short: $l^{\text{Ta}}_{\text{SD}} \sim 2$~nm \cite{Zahnd2018} --  yet are the standard materials for buffers for this type of MML stack. 

Typical TIs belonging to the commonly used Bi$_2$X$_3$ class of materials are known to form in quintuple layers weakly bound by van der Waals forces \cite{He2011, Bansal2011, Zhang2009-2}. This layering forms distinct triangular terracing at the surface of an TI surface, which raises concerns for magnetic pinning sites in TI/MML heterostructures. As such this study aims to tailor a buffer to ensure a high-quality multilayer structure and optimize for magnetic domains decoupled from the TI surface topography whilst minimising the loss of spin current between the TI and MML.  

\section{Heterostructure Growth}

\begin{figure}
    \includegraphics[width=8cm]{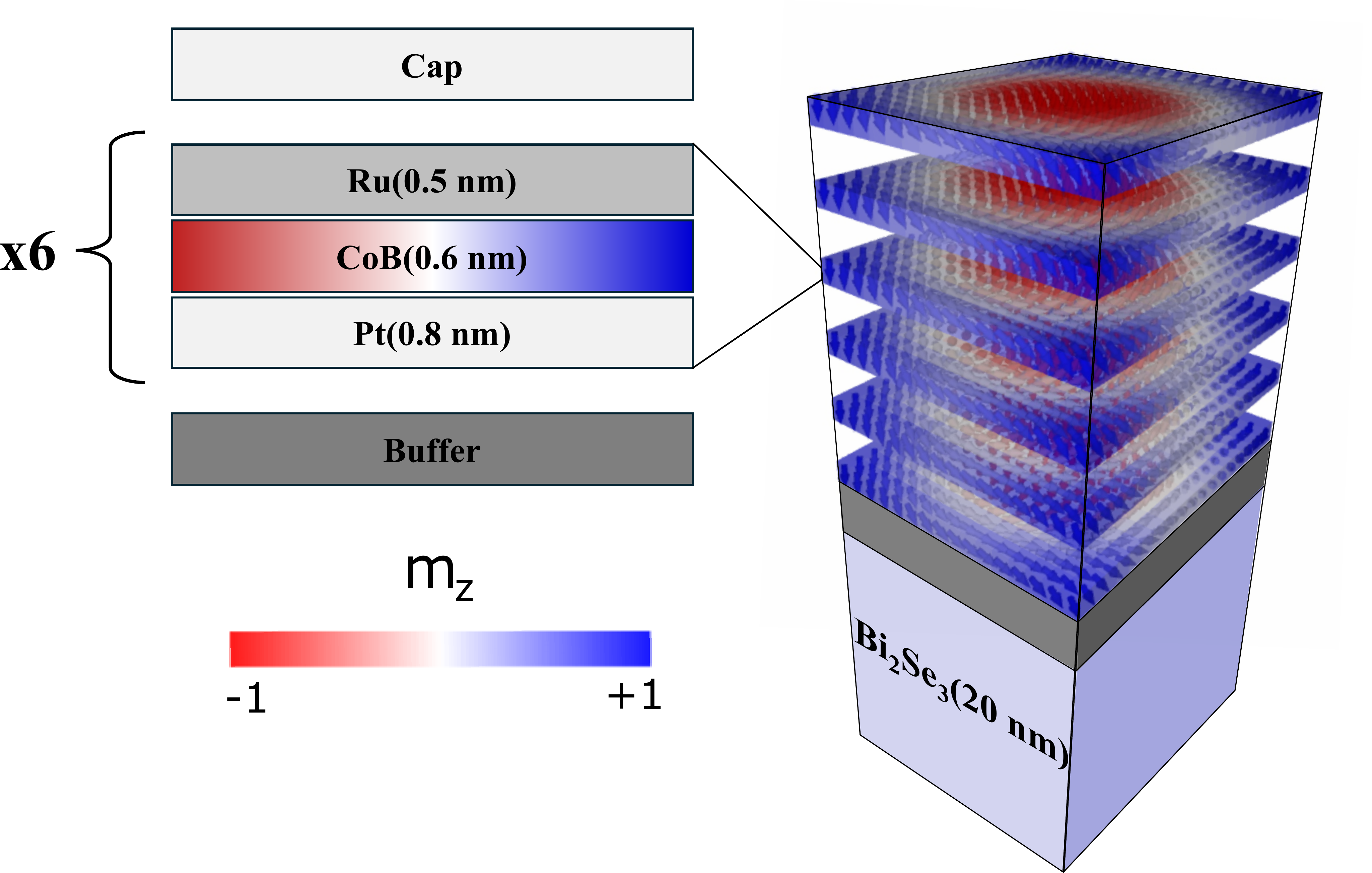}
    \caption{Schematic representing standard sample structure grown on \BS\ with illustrated skyrmion spin-texture in the CoB layers where $m_z$ lies along the out-of-plane axis.
    \label{fig:stack}}
\end{figure}

The heterostructures that we studied combine 20~nm epitaxial layers of \BS, a well-known TI \cite{Zhang2009,Xia2009,Zhang2009-2,Li2014,Wang2015}, and a [Pt(0.8 nm)/CoB(0.6 nm)/Ru(0.5 nm)]$\times 6$ MML on a Ta or Mo buffer layer as shown in Fig.~\ref{fig:stack}. The TI was grown by molecular beam epitaxy (MBE) and the multilayer was grown by dc magnetron sputtering at room temperature. 

The \BS\ (20~nm) epilayers were prepared in a solid-source molecular beam epitaxy (MBE) system that operated at a base pressure of ~7×10$^{-10} $\ mbar. The films were grown by co-depositing evaporated bismuth and selenium, both supplied from standard dual-filament Knudsen cells, on c-plane (0001) sapphire substrates. To effectively suppress the formation of selenium vacancies, the growth was carried out under Se-rich conditions, with the selenium flux at least 20 times higher than the bismuth flux. Two different samples were prepared. The first sample, characterized by narrow surface terraces $\sim$ 50 nm wide, was grown at a temperature of 235~$^\circ$C. The second, wide-terrace sample was deposited using a two-step method, with a nucleation layer of $\sim 2$ nm grown at 130~$^\circ$C and the remainder of the layer grown at 300~$^\circ$C increasing average terrace width to $\sim$ 100 nm. 

Connected MBE and sputtering chambers allowed the entire heterostructure to be deposited without breaking ultra-high vacuum, thereby minimizing degradation of the TI surface and keeping the interface between the materials as clean as possible. After transfer to the sputtering chamber, any buffer layer, followed by the MML with a 4.0~nm Pt capping layer, was deposited at room temperature. Here we have studied the buffer layer materials Ta and Mo, a $5d$ and $4d$ refractory metal, respectively.

On reference MML samples grown on thermally oxidized Si(100)/SiO$_2$(100 nm) substrates and 5.0~nm Ta buffers, PMA could be reliably obtained down to a minimum nominal Pt thickness of 0.8~nm - however the PMA strength is noticeably reduced when reducing Pt thickness beneath 1~nm \cite{Nguyen2016,Sanchez2014,Nakayama2012,Zhang2013}. To maximise the effects of interfacial PMA against shape anisotropy, the FM CoB layer was thinned to the minimum thickness to be reliably continuous, which was found here to be 0.6~nm. This optimization yields a multilayer structure with strong PMA through minimized use of heavy metals between the FM and TI layers. Samples which were not required for imaging were grown with an additional 4~nm Pt cap to prevent oxidation.

\section{Multilayer Structure on Terraced B\lowercase{i}$_2$S\lowercase{e}$_3$}

\begin{figure}
    \includegraphics[width=8cm]{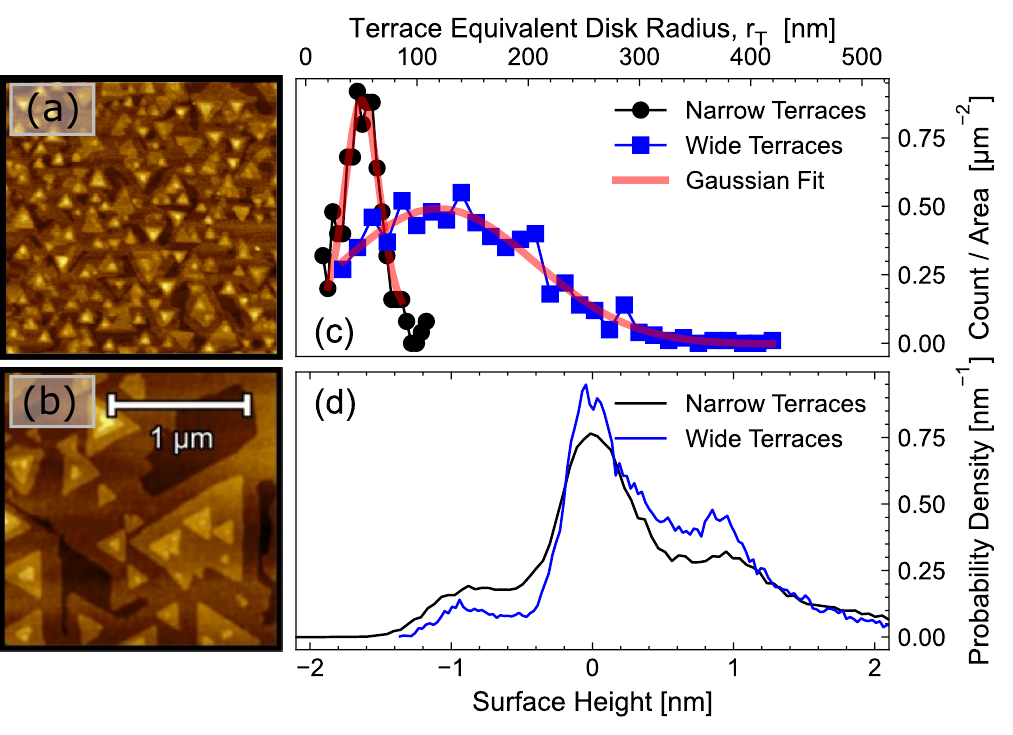}
    \caption{2~$\upmu$m~$\times$~2~$\upmu$m AFM images of \BS/Ta/MML samples grown with recipes leading to different terrace widths: narrow (a) and wide (b). The terrace widths are detected using watershed grain detection,  the areas were converted into equivalent disk radii $r_\mathrm{T}$ and fitted with Gaussian distributions shown in (c). Resultant distributions in sample height is shown in (d).}
    \label{fig:Grains}
\end{figure}

Epilayers of \BS\ grow with a terraced surface as a result of the lateral expansion and step flow growth of quintuple layers from defects during MBE growth. The terracing represents a unique challenge for \BS-based devices featuring such thin layers since a single quintuple layer, determined to be $0.96\pm0.03$~nm from X-ray diffraction (XRD) measurements, is equivalent to half a single trilayer thickness of the MML (1.9~nm).  As such the average terrace width is an important parameter to optimise for in film growth: wider terraces will mean fewer potentially disruptive terrace edges. 

\begin{figure*}
    \includegraphics[width=16.5cm]{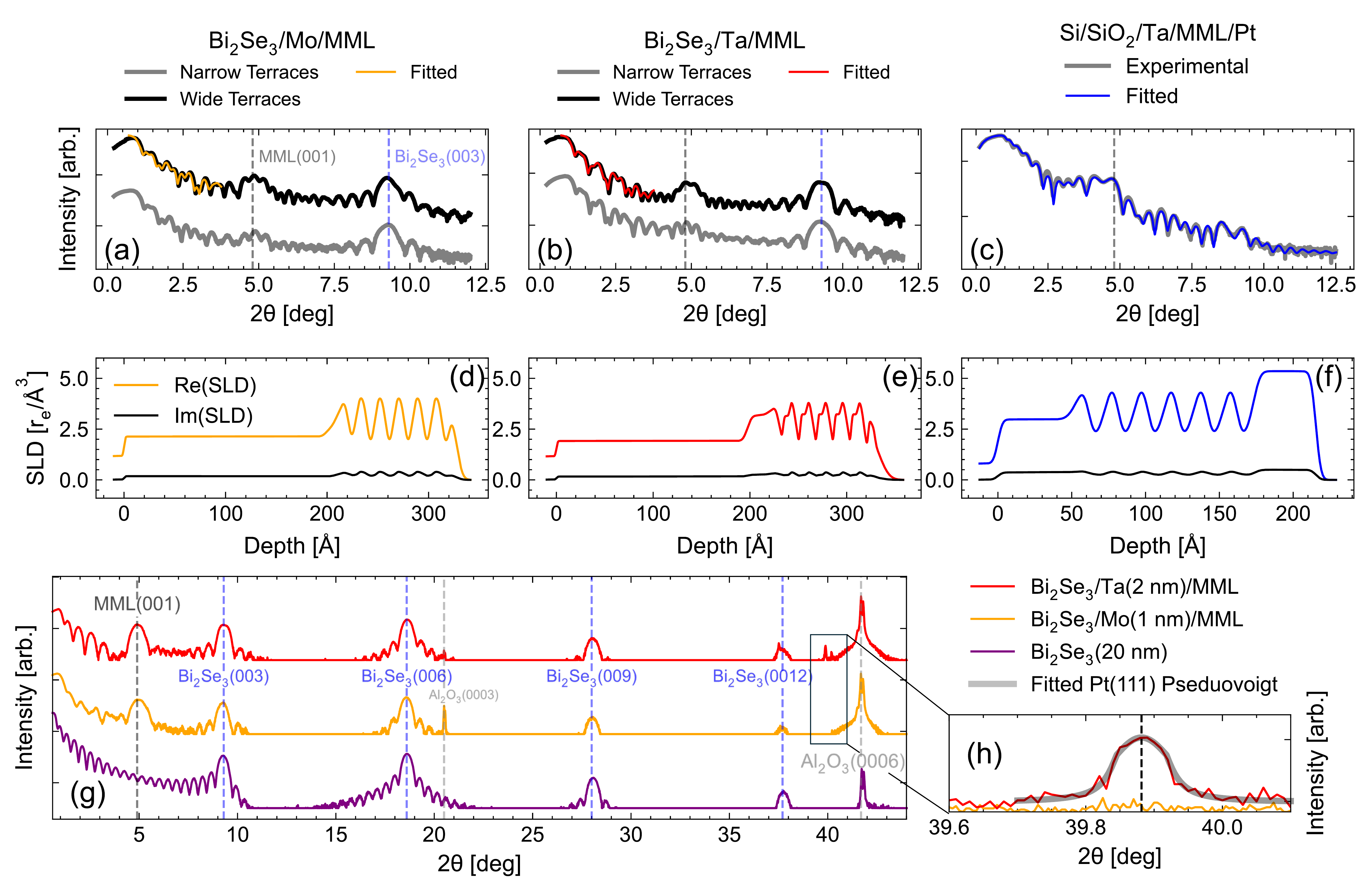}
    \caption{X-ray characterization of sample structure. (a-c) XRR curves of the main [Pt(0.8 nm)/CoB(0.6 nm)/Ru(0.5 nm)]$_{\times 6}$ MML stack as grown on \BS\ with (a) Mo and (b) Ta buffer layers as well as (c) on a thermally oxidized silicon substrate with 5.0~nm Ta buffer and 4.0~nm Pt cap. Multilayer and \BS\ peaks are marked with grey and blue dashed lines, respectively. (a-b) show curves using narrow and wide terraced \BS\ however only the wide terraced samples are fitted. The range over which the slab model is fitted is shown in a solid line.
    Axes beneath (d-f) show the fitted SLD distribution corresponding to these fits. The underlying slab models from (d) and (e) are shared with those later shown in Figs \ref{fig:PNR_example} \& \ref{fig:PNR}. (g) XRD patterns of wide-terrace \BS\ grown both with and without overlying MMLs using both buffers. Relevant peaks are labelled. (h) Zoomed axes of (g) showing the Pt $(111)$ peak. All intensity scales in this figure are logarithmic.
    \label{fig:X-rays}}
\end{figure*}

From the atomic force microscopy (AFM) of complete \BS/Ta/MML heterostructures shown in Fig.~\ref{fig:Grains}(a-b), the \BS\ surface topography can be seen to imprint through the overlying multilayer. The result of different \BS\ growth recipes on average island size and terrace width is evident. The terrace widths were detected using watershed grain detection,  using the open-source software Gwyddion \cite{Necas2012}. The distribution in equivalent disk radii, $r_\mathrm{T}$ where $\pi r^{2}_\mathrm{T} = A_\mathrm{T}$ for terrace area $A_\mathrm{T}$ can be shown to be well approximated by Gaussian distributions with mean radii of $\langle r_\mathrm{T} \rangle = 51 \pm 1$ and $120 \pm 4$~nm  for narrow and wide terraced \BS\, respectively as shown in Fig.~\ref{fig:Grains}(c). The fitted standard deviation of distributions is higher for the wide terrace recipe at 100~nm versus 20~nm for narrower terraces indicating a greater variety of island sizes and terrace widths. The terracing is evident in the sample height distributions in Fig.~\ref{fig:Grains}(d) with most of both sample surfaces being found within $\pm 1$ terrace steps of the dominant terrace height. 

Through slab-model fitting of X-ray reflectometry (XRR) measurements, our MMLs exhibit monolayer-level precision with interfacial roughnesses on the order of 2-3~\AA\ in corresponding fitted slab models shown in Fig~\ref{fig:X-rays}(a-f). Measurements were taken using a Cu K$\alpha$ source of $\lambda = 1.541$~\AA\  with a coherence length of $L_\mathrm{c} \sim {\lambda^2}/{\Delta\lambda} \approx 300$~nm. This characteristic lengthscale is sufficiently similar to the \BS\ terracing for the effect of widening the average terrace width to be apparent on the reflectivity. The extracted multilayer periodicity of 1.9 nm corresponds to a multilayer Bragg peak at $2\theta \sim 4.8^{\circ}$.
For the narrow terraces where $\left< r_\mathrm{T} \right> \ll L_\mathrm{c}$, the effective multilayer scattering length density (SLD) is averaged over multiple terraces and so multilayer features are suppressed. When the terraces become wider such that $\left< r_\mathrm{T} \right> \lesssim L_\mathrm{c}$, the effective medium approximation (EMA) begins to break down and coherent reflection from wider terraces allows for a broad characteristic Bragg peak to emerge. This is reproducible on both Mo and Ta buffer layers shown in Fig. \ref{fig:X-rays}(a-b) and confirms that the trilayer structure grown on \BS\ is resolvable. This dependence of the multilayer Bragg peak on the average terrace width is attributed to difficulties fitting high-angle reflectivity data for samples grown on the \BS.
This is addition to the Laue oscillations of the lowest order \BS\ XRD peak, which appears at $2\theta = 9.3^{\circ}$. Hence these XRR curves are only fitted in the low angle regime. 
\begin{figure*}
    \includegraphics[width=17cm]{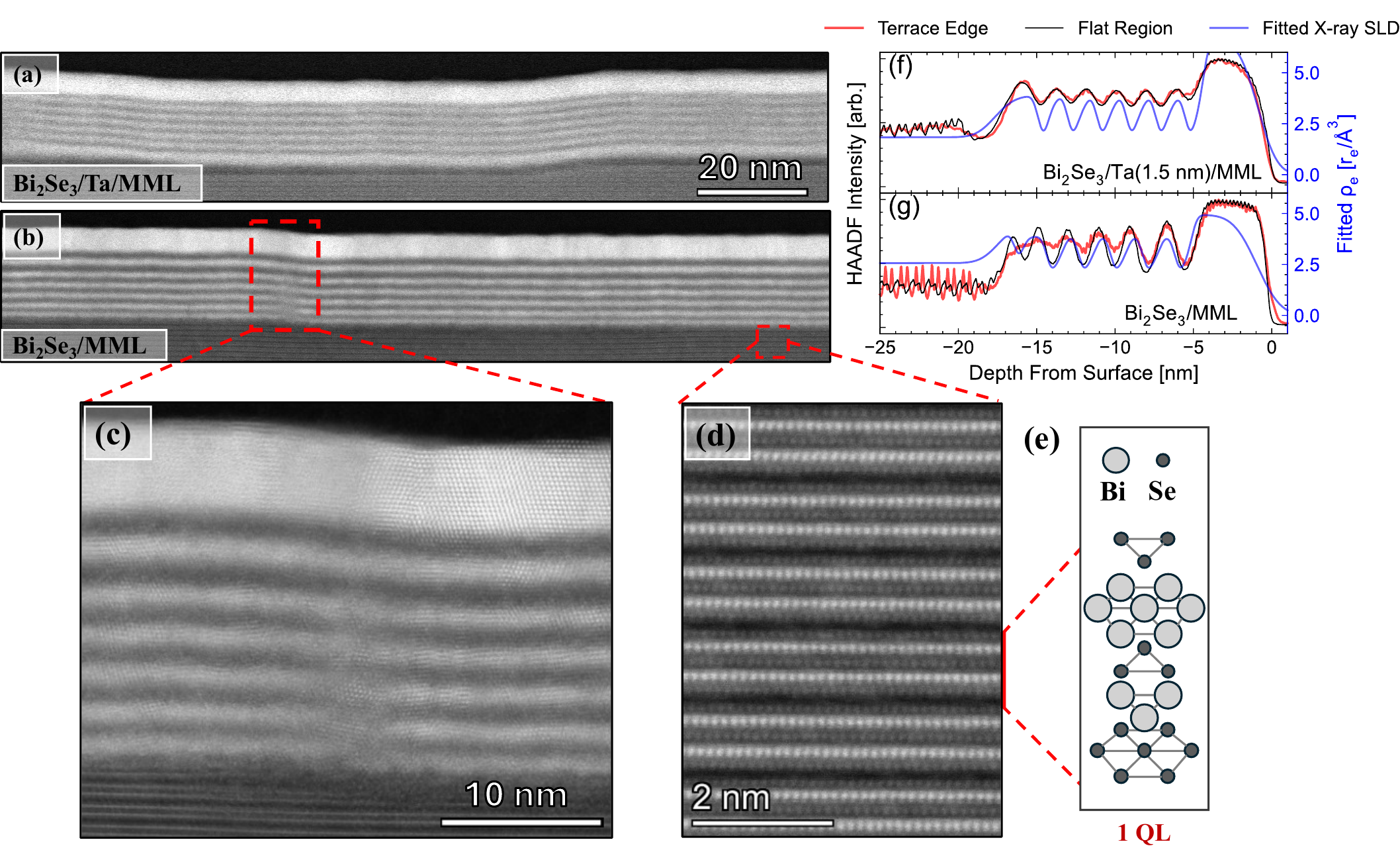}
    \caption{
    Cross sections prepared by FIB of \BS/buffer/MML/Pt(4.0 nm) heterostructures as seen under HAADF-STEM imaging. (a-b) Cross sections showing MML overlaying multiple terrace edges (a) with and (b) without a Ta (1.5~nm) buffer on narrow terrace \BS; (c)  Zoomed in section of (b) showing terrace edge with resultant layer intermixing; (d) observed quintuple layering of \BS\ with corresponding schematic of a single quintuple layer \BS\ shown in (e); (f-g) integrated line profiles of HAADF grey count at and far away from the terrace edge with (f) and without (g) buffer. Overlaid in (f-g) are the respective, co-fitted X-ray SLDs corresponding to models used in Fig. \ref{fig:PNR}.
    \label{fig:terraces}}
\end{figure*}

The full XRD patterns shown in Fig.~\ref{fig:X-rays}(g) show distinct diffraction peaks matching the \BS\ crystal structure. The large number of Laue oscillations are seen indicating good sample quality. Notably, these are seen before and after MML deposition, indicating little to no damage to the TI throughout the sputtering process. The choice between Mo and Ta buffers creates a noticeable difference in the XRD spectra. When grown without the 4 nm Pt capping layer, multilayers grown upon a Ta (2.0~nm) buffer display a diffraction peak seen at 39.88$^{\circ}$, characteristic of Pt $(111)$ ordering within the multilayer structure as shown in Fig.~\ref{fig:X-rays}(h). Pt grown on Mo (1.0~nm) does not show evidence of such Pt(111) ordering in XRD above the noise floor. This is potentially supported by subtle differences in X-ray SLD distributions where the fitted \BS/Ta/MML profile is the only sample to show three distinct metallic layers in the multilayer repeat unit, indicating sharper interfaces.

We can directly observe the structural effect of TI surface terracing on the MMLs using high angle annular dark field scanning transmission electron microscopy (HAADF-STEM), shown in Fig.~\ref{fig:terraces}. The multilayer structure over the terrace edge is observed both (a) with and (b) without a Ta (1.5~nm) buffer, and in both cases can be seen to closely follow the profile of the underlying \BS surface. The quintuple layering of \BS\ which leads to terracing is evident under closer imaging as shown in (d). Notably no Se vacancies are found on the atomic scale. Taking an averaged vertical line profile, the layers in both samples can be seen to be well defined when away from terrace edges shown in Fig.~\ref{fig:terraces}(f-g). The features in the HAADF-count broadly match the fitted X-ray SLD distributions used throughout. Some slight modulation in stack thickness is observed in the non-buffered sampled in Fig.~\ref{fig:terraces}(f) with layers being thicker towards the top of the stack, however this is attributed to the sputter targets heating up during growth leading to growth rate drift. 

At the terrace edge, the multilayer without any buffer is seen to break down due to increased layer-to-layer intermixing which worsens closer to the \BS\ interface as shown in Fig.~\ref{fig:terraces}(c) \& (g). When the Ta buffer is inserted however, this effect is not seen and the line profile is unchanged at the terrace edge with all layers remaining resolvable as shown in Fig.~\ref{fig:terraces}(a) \& (f). This continuous multilayer structure can be expected to reduce the effect of pinning of domain walls and topological spin textures at terrace edges, essential for any future device applications.

\section{Achieving Full Perpendicular Magnetic Anisotropy via Buffer Insertion}

\subsection{Magnetometry}

\begin{figure*}
    \includegraphics[width = 17cm]{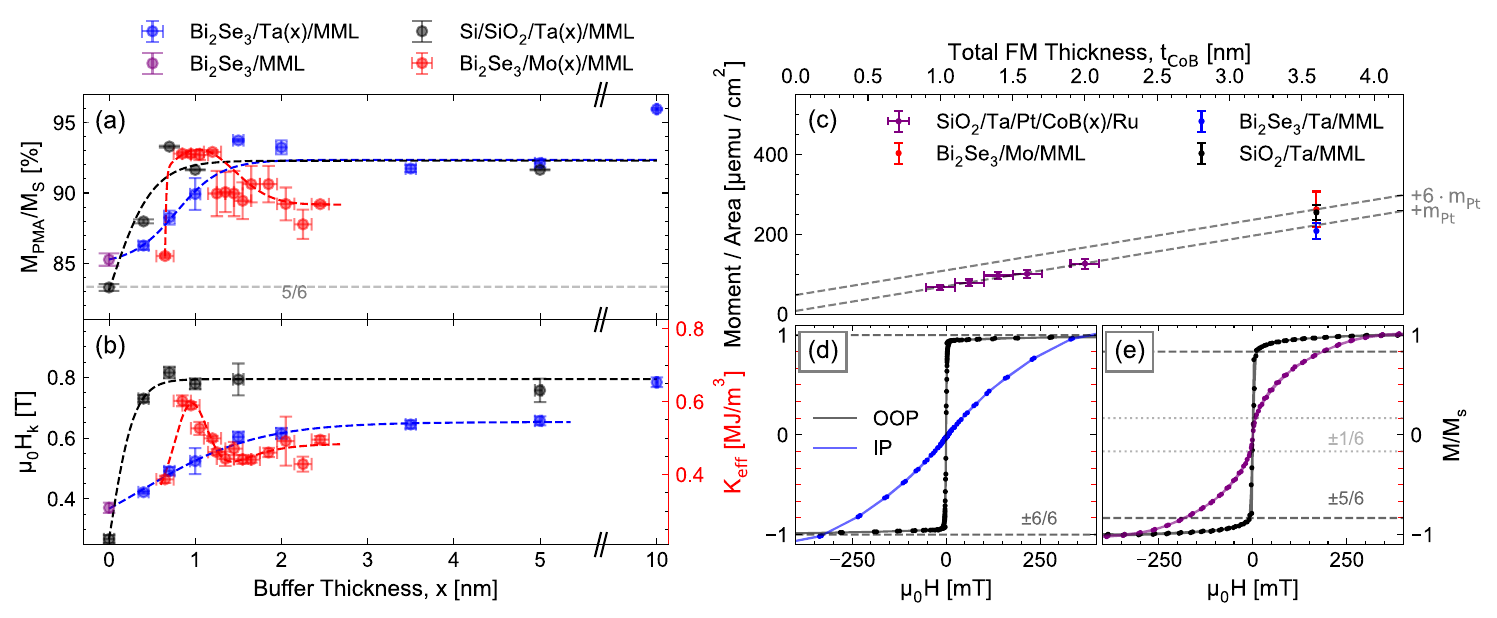}
    \caption{Buffer thickness dependence of MML magnetic properties, extracted from SQUID-VSM measurements for Ta and Mo buffer thickness series on \BS: (a) PMA fraction taken as the fraction of M$_{\text s}$; (b) the effective anisotropy field, $\mu_0 H_{\text K}$ or corresponding \keff\, assuming constant $M_\mathrm{s}$. Both (a) \& (b) include fitted sigmoid + Gaussian functions included to highlight the trend shape as well as a reference series on Si/SiO$_2$/Ta(x). (c) Moment per unit area versus total $t_{\text{CoB}}$ in the stack. A linear relation is fitted to simple Pt/CoB(x)/Ru trilayers to calculate $M_\mathrm{s} = 600 \pm 20$~emu/cc; a second is included to account for 6 proximity magnetized Pt interfaces in the MML stacks. Included are the averaged $M_\mathrm{s}$ values for Mo and Ta buffer thickness series shown in (a-b); (d-e) SQUID-VSM loops MMLs on \BS\ on a Ta(10 nm) buffer (d) or no buffer (e) taken in both IP and OOP geometries. Easily magnetizable fractions in both geometries are highlighted corresponding to a signal equivalent to an integer number of layers magnetized ($n/6$).
    \label{fig:Keff}}
\end{figure*}

To determine the magnetic properties, the samples were measured by superconducting quantum interference device vibrating sample magnetometry (SQUID-VSM) from which key magnetic parameters such as the stack-averaged, effective anisotropy field $H_\text{K}$, the here-named ``PMA ratio'', and saturation magnetization $M_\text{s}$ - could be extracted for each sample. The anisotropy field is defined as $H_\text{K} = 2 K_\text{eff}/ \mu_0 M_\text{s}$ and where $K_\text{eff} > 0$ is corresponds to an out-of-plane easy axis. The stack-averaged effective anisotropy $K_\text{eff}$ was calculated by integrating the difference between in-plane and out-of-plane $M(H)$ curves. We define the PMA ratio as $M_{\text{PMA}}/M_{\text s}$, the magnetization fraction of the sample magnetized under a small 100~Oe bias field out-of-plane as a fraction of the total saturation magnetization $M_\mathrm{s}$. This small field is needed to saturate out any domains that exist at zero field, and is used in place of the more usual squareness ratio which would be near-zero for samples with wasp-waisted hysteresis loops, despite possessing strong PMA.

As shown in Fig.~\ref{fig:Keff}(f), samples grown without a buffer between the TI and MML show easily magnetisable fractions of 5/6 and 1/6 for the out-of-plane and in-plane directions, respectively, as well as a reduced $H_\text{K}$. This suggests that five of the six CoB layers in the MML possess PMA, but one does not and is instead in-plane magnetized at low fields. Samples were then grown with a range of Mo and Ta buffer layers thicknesses $x$ to find the minimum thickness needed to recover full PMA in every layer. \keff, the PMA fraction and $M_s$ were then plotted as a function of buffer material and thickness. These values are shown in Fig.~\ref{fig:Keff}(a), (b), and (c) respectively.

Inserting a Ta buffer between the TI and the multilayer recovers the PMA fraction to $>90$\% once a thickness of $x \sim 1.5$~nm is exceeded, as shown in Fig.~\ref{fig:Keff}(a), with $H_\text{K}$ continuing to rise up to a thickness of 2~nm where $H_\mathrm{K}$ reaches a plateau at $\mu_0 H_\mathrm{K} = 0.65 \pm 0.02$~T. This is a similar process to that of Ta on natively oxidized Si, where the PMA fraction increases from $\sim 5/6$ without any buffer, plateauing once a critical thickness is surpassed. Differences to note are that on Si, the PMA ratio and $H_\mathrm{K}$ recover faster, plateauing at the same point at $x \sim 0.8$ nm and to a higher $H_\mathrm{K}$ value of $\mu_0 H_\mathrm{K} = 0.80 \pm 0.01 $~T or the equivalent of $K_{\text{eff}} = 0.66 \pm 0.01$~MJ/m$^3$. To obtain a $H_\text{K}$ value comparable to multilayers grown on natively oxidized Si/SiO$_2$ on \BS, the buffer must be increased further, seen in Fig.~\ref{fig:Keff}(b), to agree with Si at $x = 10$~nm. At this large thickness of Ta we can expect that the multilayer would be effectively shielded from any spin-signal from the \BS\, leaving it redundant.  

Repeating the same experiment with Mo instead of Ta, a different behavior was observed, with the both $H_\text{K}$ and the PMA ratio peaking at a low thickness of $0.94 \pm 0.06$~nm to an increased $\mu_0 H_\mathrm{K} = 0.72 \pm 0.05$ T. Increasing the buffer thickness further leads to a sharp decrease in $H_\text{K}$ and small decrease in the PMA ratio. This shows there is sufficient Pt $(111)$/Co hybridization for PMA despite the Pt $(111)$ peak not being visible in XRD measurements in Fig.~\ref{fig:X-rays}(d). We can hypothesise that there is increased intermixing between Pt and CoB when grown on Mo versus Ta leading to insufficient long range ordering for the Pt $(111)$ peak to be visible. The drop observed in $H_\mathrm{K}$ could be explained as this effect worsening with increased Mo thickness.  

We can therefore conclude that depositing at least 1.5~nm of Ta or $\sim 0.9$~nm of Mo on the \BS\ epilayer surface as a buffer layer recovers the PMA in the layer that is in-plane magnetized with no buffer. The choice of buffer comes with  additional subtle changes in the MML's magnetic characteristics.

Multilayers grown on both Ta and Mo buffers show $M_\mathrm{s}$ values consistent with a trilayer Pt(1.5 nm)/CoB($x$)/Ru thickness series as grown on Ta (5~nm) on natively oxidized Si, as shown in Fig.~\ref{fig:Keff}(c), where $M_\mathrm{s}$ for Co$_{68}$B$_{32}$ is $600 \pm 20$~emu/cc. This indicates no substantial magnetic dead layers forming as a result of the \BS\ or of the reduced Pt thickness. Proximity induced magnetism (PIM) in Pt is indicated by a non-zero intercept of $(m/A)_{\text{Pt}} = 8 \pm 3$ $\upmu$emu/cm$^{2}$ per Pt interface. 

\subsection{Polarized neutron reflectometry}

\begin{figure}
    \includegraphics[width=8cm]{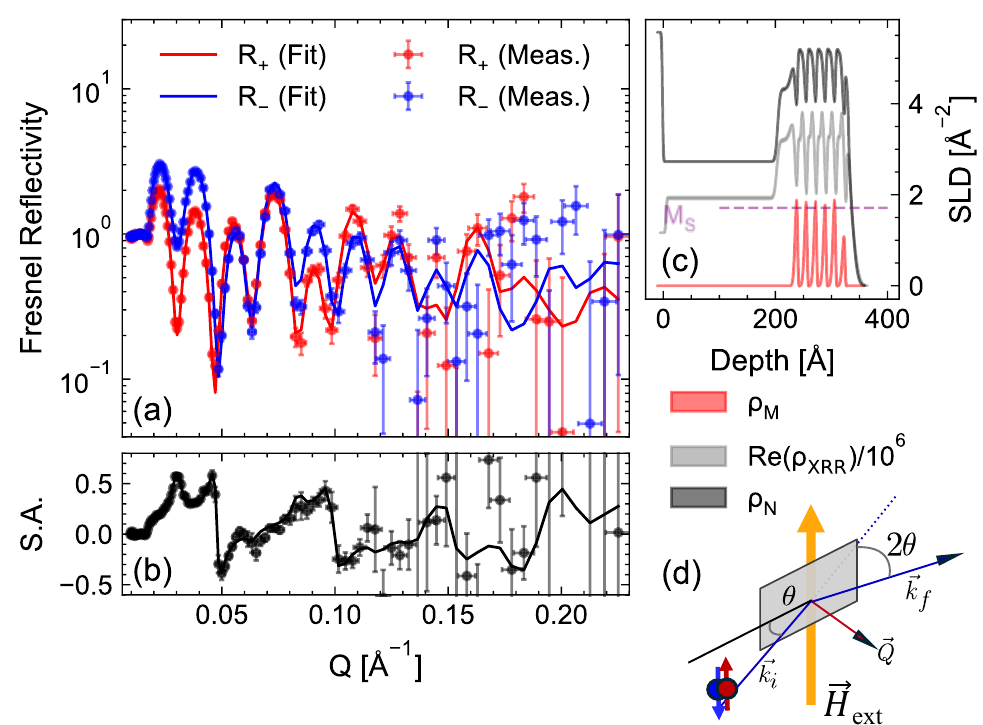}
    \caption{Example polarized neutron reflectometry data: (a) shows Fresnel reflectivity curves for both neutron polarizations: the resultant SA is plotted in (b). Simulated results from a fitted slab model are shown in (a) and (b): the corresponding model's $1 \sigma$ confidence intervals for magnetic and non-magnetic neutron scattering distributions are shown in (c) as well as the co-fitted X-ray SLD from \ref{fig:X-rays}(e). The equivalent $\rho_\mathrm{M}$ from the calculated CoB $M_\mathrm{s}$ value taken from Fig.~\ref{fig:Keff}(c) is shown as reference; (d) shows the sample geometry with respect to the neutron beam and applied field.}
    \label{fig:PNR_example}
\end{figure}

Whilst the conventional magnetometry measurements show that full PMA can be recovered in these ultrathin films on \BS, they still highlight that the magnetic anisotropy can vary greatly throughout the stack. This could be problematic when spin-torques are predominantly exerted on the bottom-most FM layer where we cannot easily directly image. A solution to this is to verify the multilayer stack is magnetically homogenous such that we can be confident magnetic textures seen through surface-sensitive measurements correspond to the full texture throughout the stack through magnetostatic coupling. 

Polarized neutron reflectivity (PNR) was used to obtain a depth resolved magnetization profile of multilayers grown on the various buffers. In this way we can identify the layer that lacks PMA when no buffer is present as well as any large changes in anisotropy. Spin-polarized neutron reflectivities $R_{+}$ and $R_{-}$  were taken as a  function of wavevector transfer $Q$ through time-of-flight detection on the POLREF instrument at ISIS, and the spin asymmetry (SA), defined as SA $ = (R_{+} - R_{-})/(R_{+} + R_{-})$, between them was calculated. An example PNR measurement is shown in Fig.~\ref{fig:PNR_example}, including the measurement geometry. SAs were measured at applied in-plane fields from full or near saturation ($\mu_0 H = 700$~mT) down to the minimum applicable field to prevent beam de-polarization ($\mu_0 H = 10$~mT). Since PNR in the geometry we used, see Fig.~\ref{fig:PNR_example}(d), is sensitive to the in-plane component of magnetization, our experiment was designed to measure that component as it increases whilst the in-plane applied field forces the magnetization into the magnetically hard film plane. When the PNR data were fitted (Fig.~\ref{fig:PNR_example}(a) \& (b)) using the Refl1D code \cite{refl1d}, the resulting magnetic SLD distribution appears as a spike for each of the six magnetic CoB layers (Fig.~\ref{fig:PNR_example}(c) with an amplitude proportional to that in-plane component. 

\begin{figure*}
    \includegraphics[width = 17cm]{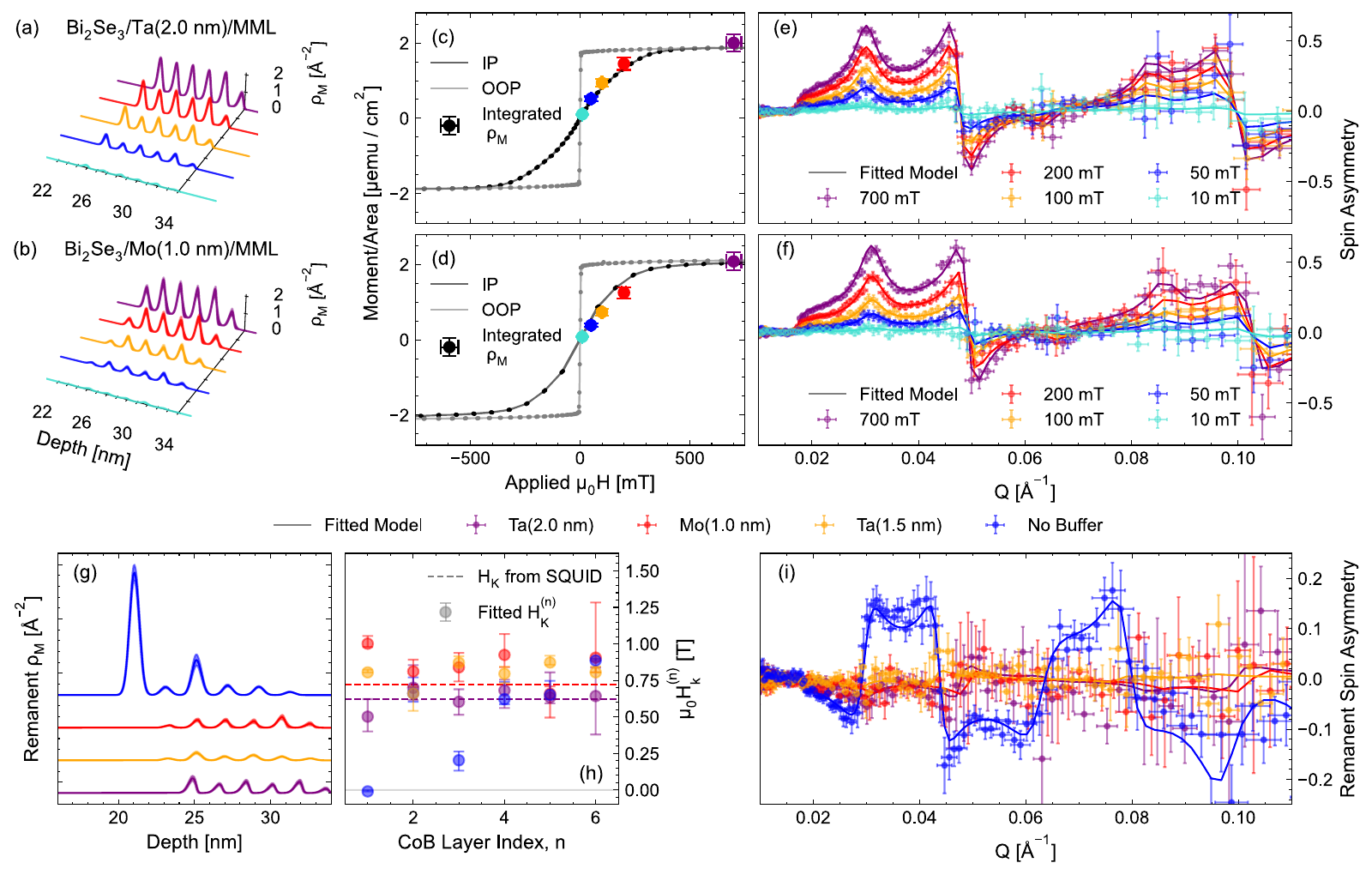}
    \caption{PNR as a function of external applied field. (a-b) Fitted magnetic scattering length density profiles, $\rho_\text{M}(z)$ for [Pt/CoB/Ru]$_{\times 6}$ multilayers grown on Ta~(2~nm) (a) and Mo~(1~nm) (b) buffer layers on wide-terrace \BS. Opaque and translucent profiles show 1 and $2 \sigma$ confidence ranges, respectively. (c-d) Corresponding magnetometry data for MMLs on Ta~(2~nm) and Mo~(1~nm) buffers respectively. Out-of-plane and in-plane SQUID-VSM measurements are shown over the integrated $\rho_\text{M}$ profiles from (a-b). (e) and (f) Fitted spin-asymmetry profiles for Ta and Mo buffers, respectively, with simulated data corresponding to the $\rho_\text{M}(z)$ profiles shown in (a-b). (g) Fitted SLD distributions for various buffers  taken at the minimum 10~mT alignment field. (h) Individual, layer-resolved $H_{\text{K},n}$ values obtained from the model fittings for various buffer types. Stack averaged $H_\text{K}$ values extracted from SQUID measurements in Fig.~\ref{eq:keff}(b) are shown for reference. (i) Fitted low field spin asymmetries corresponding to the slab models in (g).}
    \label{fig:PNR}
\end{figure*}

The PNR confirms the SQUID-VSM result that both Ta~(2~nm) and Mo~(1~nm) buffers yield multilayers with zero in-plane remanance, demonstrated by a vanishing spin-asymmetry under a 10 mT field, which shows that there is no in-plane component of magnetization present. The lack of any notable change in shape of the spin asymmetry with field, shown in Fig.~\ref{fig:PNR}(e) and (f) suggest magnetic profiles changing in magnitude and not substantially in shape. We can conclude that magnetic anisotropy is uniform throughout the multilayer stack, which possesses full PMA. 

Measured neutron reflectivity curves were fitted to nuclear and magnetic SLD profiles ($\rho_\text{N}$ and $\rho_\text{M}$) using slab model fitting. Due to the low thickness of the individual layers, neutron data was co-fitted with the X-ray data and models shown in Fig. \ref{fig:X-rays}(a-b,d-e) to extend our range in $Q$. Our magnetic slab models assume a phenomenological, Langevin-type $M(H)$ dependence and fit each CoB with a individual effective anisotropy, $K_{\text{eff}, n}$ where the magnetic SLD in the $n^\text{th}$ CoB layer follows:

\begin{equation}
    \rho_{M,n}(H) = {\mu_0}{\alpha} M_\text{s} {\cal L}(a_n\mu_0H)
\end{equation}

where ${\cal L}(a_nH)$ is the Langevin function and $\alpha = 2.32$~\AA$^{-2}$T$^{-1}$ is a conversion factor \cite{Maranville2016}.
This allows for field-dependent simulations based on a single scaling parameter $a_n$ per layer which can be related to $K_{\text{eff}, n}$ and has units of T$^{-1}$. The final fits are shown in Fig.~\ref{fig:PNR}(a-b) and were obtained using the open source library Refl1D \cite{refl1d}. Integrating $\frac{1}{\mu_0\alpha}\int \rho_M \,\text{d}z$ yields the moment per unit area which can be shown to agree with the measured SQUID-VSM data shown in Fig.~\ref{fig:PNR}(c-d).

In fitted $\rho_\text{M}$ distributions at saturation where $H_\text{K} < H_{\text{ext}}$, all but the top-most layers can be shown to agree with the $M_\text{s}$ values of CoB as shown in Fig.~\ref{fig:PNR_example}(c). Discrepancies in the top-most layer can be discounted as being due to the TI and are more likely the result of slight oxidation due to the thin cap required for subsequent synchrotron imaging (see next section) or the result of removing the overlying Pt layer.

Extending over all fields, the model achieves goodness of fit of $\chi^2 = 2.84$ and 2.87 for Ta (2.0~nm) and Mo (1.0~nm) buffers, respectively. A more detailed description of the fitting process can be found in the Supplementary Material \cite{Supp}. Integrated $\rho_\text{M}$ distributions give an estimation for the dipole moment per unit area, which is shown to closely agree with values extracted from SQUID-VSM measurements in Fig.~\ref{fig:PNR}(c) and (d). The extracted, layer-resolved anisotropy fields $H_{\text{K},n}$ shown in Fig.~\ref{fig:PNR}(h) are broadly distributed around the stack averaged values determined by SQUID-VSM in the $\mu_0 H_{\text{K},n} \sim 0.5$-1.0~T range. Averaging over these layer-resolved anisotropies for the Ta(2.0 nm) buffer yields $\frac{1}{6}\sum_{n=1}^{6}\mu_0 H_{\text{K},n} = 0.63 \pm 0.03$~T, agreeing with the corresponding value from SQUID-VSM. Likewise on the Mo (1.0~nm) buffer, this method slightly overestimates $H_K$ compared to the SQUID-VSM-derived value with $\frac{1}{6}\sum_{n=1}^{6}\mu_0 H_{\text{K},n} = 0.84 \pm 0.05$~T, however due to the maximum applied field 0.7~T~$< \mu_0H_{K}$ for the Mo (1.0~nm) buffer, unlike for the Ta(2.0 nm), this may limit the accuracy of the final fitted values.   

Upon removal of the buffer layer, the remanent signal becomes non-vanishing as seen in Fig.~\ref{fig:PNR}(i). Furthermore as the shape of the SA profile changes, we can infer the magnetic profile changed in shape compared to saturation and therefore is rendered inhomogeneous throughout the stack. Through slab-fitting shown in Fig.~\ref{fig:PNR}(g) this can be shown to be localized to a remanent, in-plane magnetization in the bottom of the multilayer stack. Here the interfacial layer's modelled $K_{\text{eff},1}$ drops to zero: consistent with this single CoB layer losing PMA and consistent with the missing 1/6 from the PMA fraction seen in Fig.~\ref{fig:Keff}(a). This is consistent with previous literature  where single films grown directly onto TIs did not posses PMA \cite{Pan2022,Dc2018}. We see here where a buffer is not used, this first interfacial layer acts as sufficient buffer material such that the following 5 layers form above with PMA.  

\section{Magnetic Domain Dependence on Buffer Selection}
\begin{figure*}
    \includegraphics[width = 17cm]{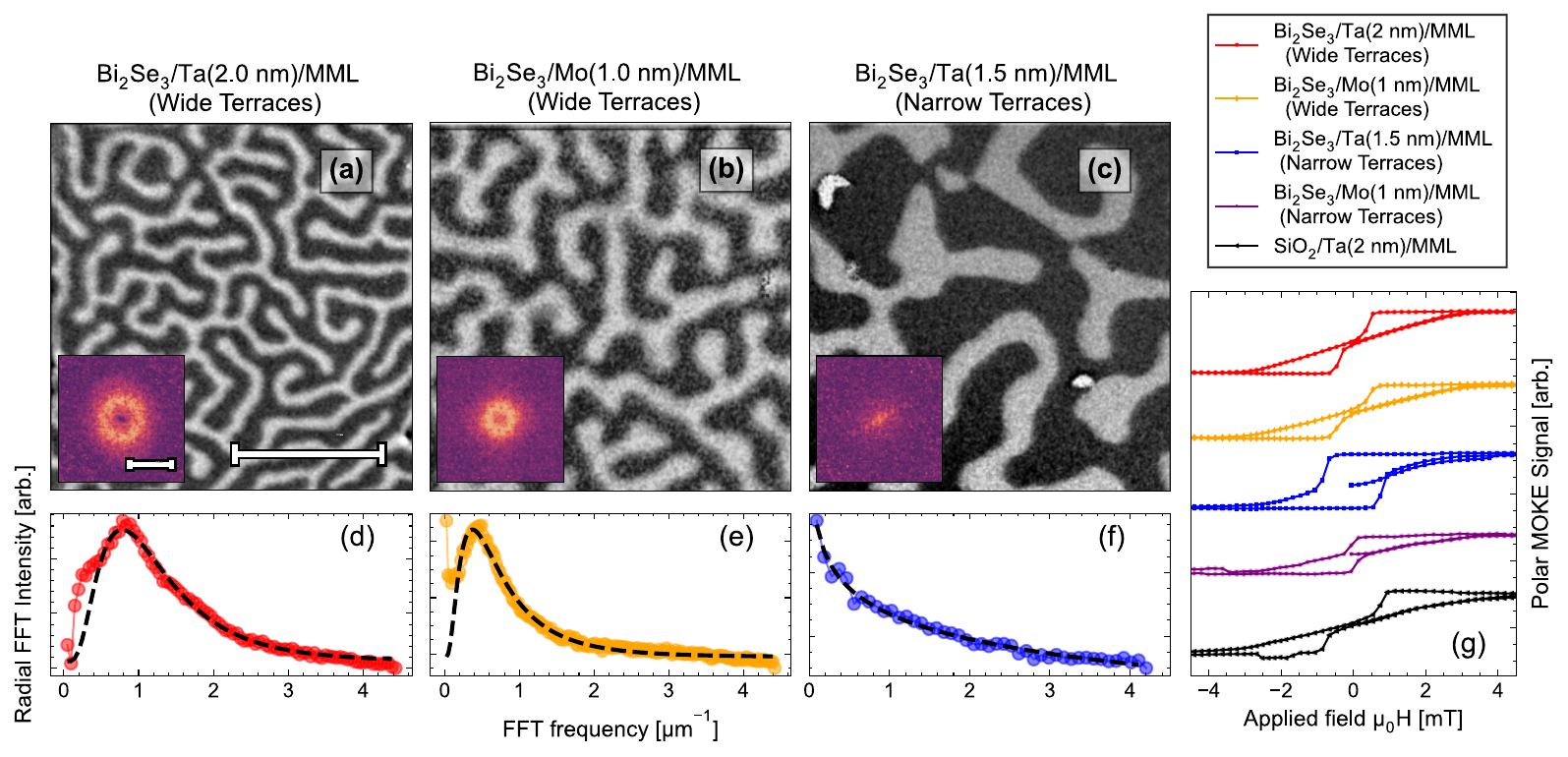}
    \caption{Imaging of domains in the MMLs. (a-c) XMCD-PEEM difference images of magnetic domains in samples with (a) Ta (2.0~nm), (b) Mo (1.0~nm), and (c) Ta (1.5~nm) buffers with either wide (a-b) or narrow (c) terraces. Scale bars represent 2$\upmu$m. (d-f) Azimuthally-integrated, radial FFT intensities with fitted log-normal distributions shown in black for the same three images. Corresponding FFTs are shown as insets. The included scale bar represents 2$\mu$m$^{-1}$. (g) Polar MOKE hysteresis of MMLs with virgin curves as grown on different buffer layers, TI surface terrace width, and compared against identical MMLs grown on thermally oxidized Si/Ta(5.0 nm).
    \label{fig:PEEM}}
\end{figure*} 

The realization of surface state-mediated spin torques on complex magnetic textures and domains will require not only tuning of the anisotropy but also of the nucleated domain structure, raising again the question of the dependence underlying topography. The effect of the average terrace width on magnetic hysteresis in polar magneto-optical Kerr effect (MOKE) measurements can be seen in Fig.~\ref{fig:PEEM}(g). Decreasing terrace width on a Mo (1.0~nm) buffer results in a loss of the linear region near zero field indicative of domain pinning increasingly disrupting the formation of demagnetization-induced labyrinthine domains, causing the loop to become squarer when the TI surface terraces are narrow. This can be similarly seen on a Ta~(1.5~nm) buffer where the OOP hysteresis becomes almost fully square when grown on narrow \BS\ terraces.

The effect on the domain structure can be seen directly using photoemission electron microscopy (PEEM) with X-ray magnetic circular dichroism (XMCD) magnetic contrast, carried out at the I06 beamline at Diamond Light Source. 

The XMCD-PEEM images, taken with the photon energy tuned to the Co L$_3$ edge, are shown in Fig.~\ref{fig:PEEM}. These are difference images, taken by subtracting and normalizing images taken with left and right handed circularly polarized X-ray photons. Photoemissive techniques are very surface-sensitive and so the signal is dominated by the top magnetic layer. As the PNR measurements show that the layer-to-layer anisotropy is rather uniform, we can assume domains do not posses substantial depth dependence and couple through standard magnetostatic effects so as to appear in every layer in the stack. This is supported by basic micromagnetic simulations using a pristine, defect-less model with values extracted from magnetometry \cite{Supp} performed using the MuMax3 code library \cite{Vansteenkiste2014}.

Our XMCD-PEEM results show that, in samples without proper optimization of the \BS\ surface morphology that hence possess terraces that are narrow, magnetic domains lack a clear periodicity when demagnetized, as shown in Fig.~\ref{fig:PEEM}(c,f), and appear highly pinned to the \BS\ surface in (c). On samples with wide TI surface terraces, MMLs on both Mo and Ta buffers demonstrate labyrinthine domain patterns, as shown in Fig.~\ref{fig:PEEM}(a-b), with well defined periodicities seen  in fast Fourier transformed (FFT) images, insets of Fig.~\ref{fig:PEEM}(d-e), and appear to be fully isotropic in-plane. Azimuthal averages of these FFTs were fitted with log-normal profiles, yielding peaks corresponding to domain widths  in real space of $\Delta^{\text{Ta}} = 600 \pm 200$ nm and $\Delta^{\text{Mo}} = 1.1 \pm 0.6~\upmu$m for Ta (2.0~nm)  and Mo (1.0~nm) buffers respectively. Identical multilayers grown on Si wafers with 100~nm thermal oxide using Ta~(5.0~nm) buffers show an average domain width of $\Delta^{\text{SiO}_{2}/\text{Ta}} = 400 \pm 100$~nm extracted from MOKE microscopy \cite{Supp}.  

These images show that growing multilayers on \BS, regardless of buffer, leads to changes in the domain structure due to the QL surface terraces, however the narrower distribution in $\Delta$ seen in the Ta~(2.0~nm) buffer compared to Mo~(1.0~nm) seems indicative of a reduced effect of pinning from the underlying \BS. As the periodicity is characteristic of competition between exchange, the DMI, the anisotropy, and magnetostatic effects the widened and less labyrinthine domain structure on the Mo~(1.0~nm) buffer may be attributed to its increased effective anisotropy and apparent differences in Pt ordering. Due to the constraining nature of the molybdenum buffer thickness \keff\, dependence, there is little room available for optimization by altering Mo buffer thickness. Despite this, as domain structure is greatly improved on the Mo on wide terraces compared to Ta on narrow terraces, we conclude that the \BS\ optimization is the greater factor in domain optimization than the buffer choice. 

\section{Conclusions}

The insertion of a thin refractory metal buffer between a \BS\ epilayer and a subsequently deposited MML has been shown to recover full stack-averaged PMA for multilayers grown in the ultra-thin limit for suitably chosen buffer thicknesses. We achieve a similar degree of sample quality of magnetic films as those grown on standard Si/SiO$_2$ substrates using Ta and Mo buffer layers, highlighting their different thickness dependences, and indicating with PNR that these multilayers are magnetically uniform. This PMA is lost for the first magnetic layer if this buffer is omitted, and HAADF-STEM imaging showing increased intermixing at terrace sites leads to a local-breakdown in multilayer structure rendering the system unsuitable for any potential SOT-driven spin-texture motion. Inserting a buffer layer is shown to ensure the multilayer remains continuous and well defined as it overlays the \BS\ epilayer surface topography.

Through optimization of \BS\ growth to control the surface morphology of the TI epilayer, we can achieve a labyrinthine domain ground state at zero field - promising for hosting skyrmion textures. Whilst the main variable to control appears to be the average \BS\ surface terrace width, the choice of buffer is additionally shown to result in changes to the domain structure. We have shown that most well-defined domain ordering is obtained when the MML is grown on Ta~(2.0~nm) as opposed to the lighter, thinner Mo~(1.0~nm) buffer resulting in some trade-off between terrace-induced pinning and transparency to the spin-current generated in the TI. 

We conclude that the SOT-driven spin texture manipulation, now well established in heavy metal/multilayer films, may well be achievable using the Bi$_2$X$_3$ class of topological insulators as the SOT source through correct buffering. To this end both tantalum and molybdenum show promise for use in such devices so long as their respective strengths are considered.

\begin{acknowledgments}
S.H., A.Y., S.C., S.S., T.A.M. and C.H.M. acknowledge support from the EPSRC Programme Grant `CAMIE' (EP/X027074/1) and 'NAME' (EP/V001914/1).
The samples were grown in the Royce Deposition system at the University of Leeds, which is supported by the Henry Royce Institute, United Kingdom, through Grants No. EP/P022464/1 and No. EP/R00661X/1. Experiments at the ISIS Neutron and Muon Source were supported by beamtime allocations RB2410236 \& RB2510555 from the Science and Technology Facilities Council. XMCD-PEEM imaging was undertaken at Diamond Light Source under sessions MM37770-3 and MM38770-1.

\end{acknowledgments}

\section*{Data Availability}

The data associated with this paper are openly available at the Research Data Leeds repository~\cite{data}. Raw data associated with the ISIS experiments are available at the following URLs: \url{https://doi.org/10.5286/ISIS.E.RB2410236} and \url{https://doi.org/10.5286/ISIS.E.RB2510555}.

\bibliography{library.bib}

\end{document}


\maketitle
\thispagestyle{empty}
\doublespacing
\section{Measurements}
\subsection{X-ray Reflectivity Measurements}
X-ray experiments were performed using a Bruker D8 discovery X-ray diffractometer with a Cu K-$\alpha$ X-ray source. Samples were aligned for specular reflection for XRR at $2\theta = 1^{\circ}$ grazing incidence and for XRD measurements using the $2\theta = 41.69^{\circ}$ Al$_2$O$_3$ substrate diffraction peak. Both measurements were performed in the standard $\theta/2\theta$ geometry. SLD distributions on silicon were simulated and fitted to the XRR data using the GenX software package \cite{Glavic2022}. Where X-rays were co-ffitted with neutrons, the Refl1D python library was used \cite{refl1d}.

\subsection{High-Angle Annular Dark Field STEM (HAADF STEM)}

The cross-section lamella for STEM imaging was prepared using a dual-beam focused ion beam integrated scanning electron microscope (Thermo-Fisher Scientific FEI G5 CX). After electron-beam deposition of C and Pt on the surface, the samples were thinned to electron transparency. The STEM imaging was carried out using a Thermo-Fisher Scientific double-tilt STEM holder in the Thermo-Fisher Scientific FEI probe-corrected monochromated XFEG Spectra 300. The microscope was operated at 300 kV. The convergence angle was 30 mrad, and the collection angle for ADF images was 52–200 mrad using the HAADF detector. All STEM images were processed using Thermo Fisher Scientific drift corrected frame integration Velox software to correct for drift/scan distortion.

\subsection{SQUID-VSM and MOKE Microscopy}

All measurements were conducted on a Quantum Design MPMS-3 SQUID-VSM magnetometer. Samples were demagnetized and measured through -1 to 1~T forward and reverse field sweeps. SQUID signals were corrected by subtracting a linear diamagnetic background from Al$_2$O$_3$/Bi$_2$Se$_3$, correcting for trapped flux in the superconducting coils by correcting $H$ using a Pd reference sample of known susceptibility and dividing by an IP or OOP filling factor. Effective anisotropies were calculated by calculating the difference in out-of-plane and in-plane, $M(H)$ integrals with respect to $H$.

\begin{figure}
    \centering
    \includegraphics[width=0.6\linewidth]{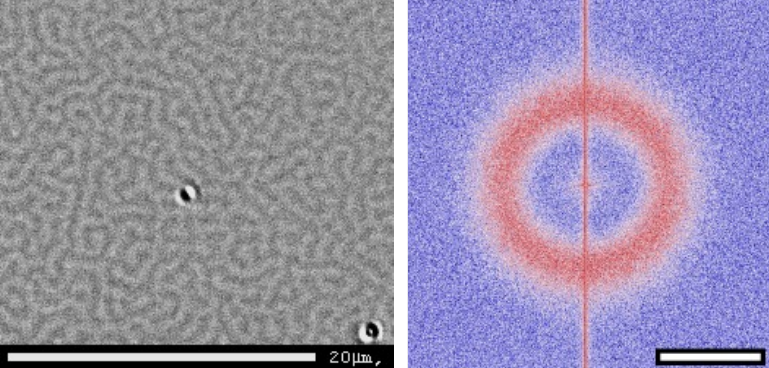}
    \caption{Polar MOKE image of Si/SiO$_2$/Ta(5.0 nm)/MML reference sample. FFT is shown inset with scale bar corresponding to 1$\mu$m$^{-1}$}
    \label{fig:MOKE}
\end{figure}
All MOKE measurements were conducted in the polar configuration for OOP magnetic sensitivity on a Kerr microscope with white LED source. Signals were corrected for Faraday components and thermal drift of the LED. The magnetic field was applied using an air core coil to avoid magnetic remanance from a permanent pole piece, and samples were measured up to a maximum applied field of $\pm 30$~mT. Domains were observable being in the lower limit of what is optically resolvable as shown in Fig. \ref{fig:MOKE}. A domain periodicity on the Si reference sample of $\lambda_{\text{FFT}/2} \sim 600$ nm was observed.

\subsection{X-ray Photoemission Electron Microscopy}

All XMCD X-PEEM measurements were conducted at the I06 beamline at the Diamond Light Source. Images were collected at the Co L$_3$ edge at room temperature and under a 15~kV bias. These were taken with a thin 1 nm Ru cap shortly after growing to maximise counts from photoemission. Little/no evidence of peak-splitting or shouldering implies sample remains unoxidised despite the thin cap. Multiple images using both X-ray helicities were taken, subtracted, drift-corrected and averaged over. 
\begin{figure}
    \centering
    \includegraphics[width=0.5\linewidth]{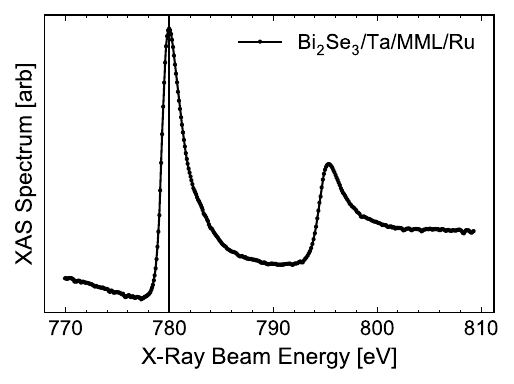}
    \caption{Example Co XAS pattern of multilayer on Bi$_2$Se$_3$ with 1 nm Ru cap. Line shows alignment to Co L$_3$ edge.}
    \label{fig:XAS}
\end{figure}
X-ray absorption spectroscopy (XAS) measurements imply minimal oxidation of Co under the 1~nm Ru capping due to a lack of any noticeable shoulder appearing on the peaks in the XAS spectra as shown in Fig. \ref{fig:XAS}. The vanishing total XMCD signal under zero applied field is consistent with maze domain formation with equal areas of up and down domains.

\section{Polarized Neutron Reflectometry}
\subsection{Neutron Measurements}
All PNR measurements were conducted on the POLREF reflectometer at the ISIS Neutron and Muon Source. Measurements are time of flight, polarised in-plane and taken in specular geometries. Reflected intensities in neutron channels $R_{+}$ and $R_{-}$ were collected.

\begin{figure}
    \centering
    \includegraphics[width=0.6\linewidth]{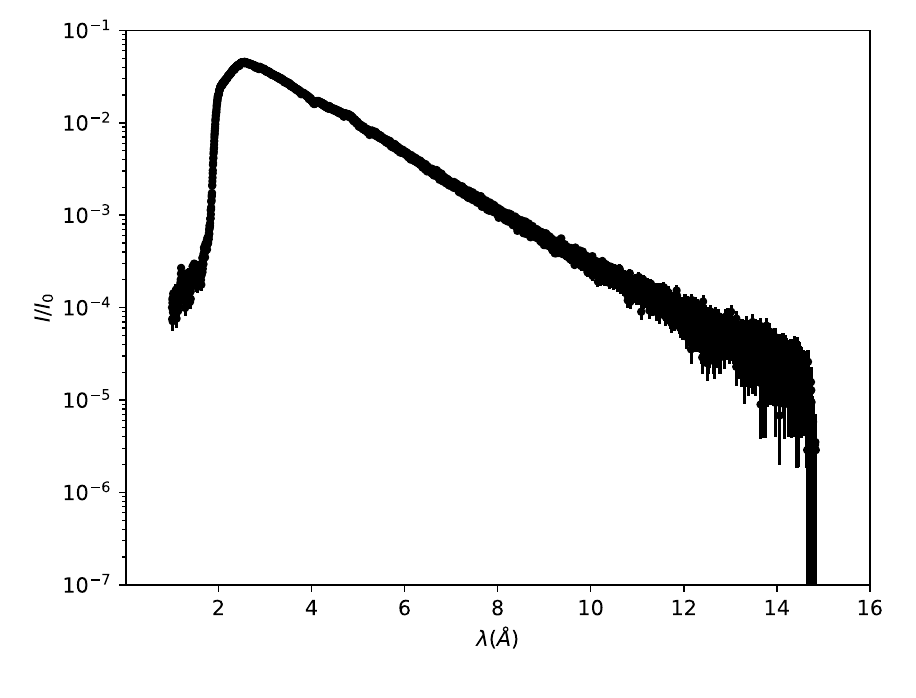}
    \caption{Transmission statistics used for data reduction showing distribution of time-of-flight detected neutrons.}
    \label{fig:transmission}
\end{figure}

Each reflectivity curve involved measurements at $2\theta = 0.1^{\circ}, 1.2^{\circ}, 2.3^{\circ}$ at 1, 2 and 8 hours respectively. To ensure good counting statistics samples were prepared as 2 inch diameter wafers. The in-plane external magnetic field was applied using a room-temperature magnet from 0.7~T decreasing to near remanance. Measurements must be conducted at a minimum of 10~mT or higher to maintain reliable beam polarisation. Care was taken to obtain good transmission statistics over mutliple hours as shown in Fig. \ref{fig:transmission}.

\begin{figure}
    \centering
    \includegraphics[width=0.9\linewidth]{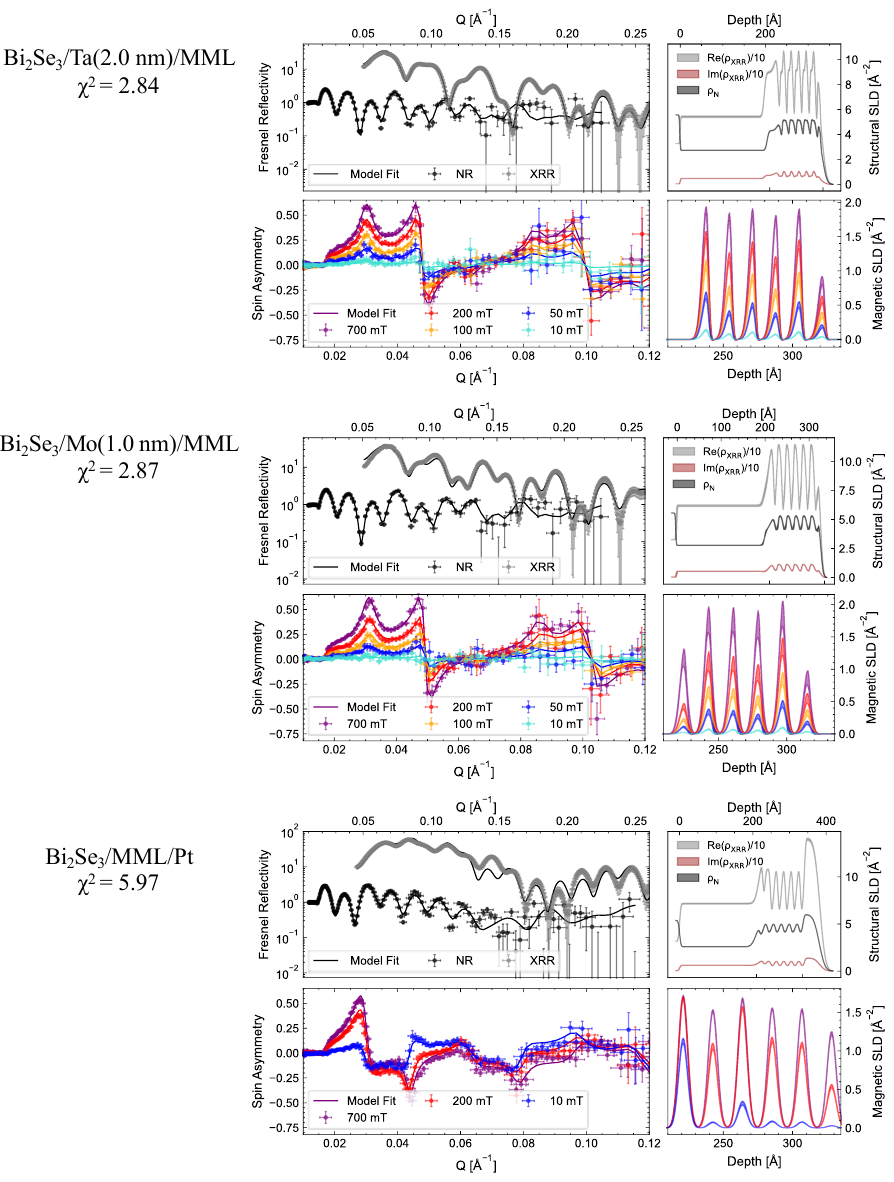}
    \caption{PNR \& XRR  curves fitted by a shared multilayer slab model for, from top to bottom, a Ta(2.0 nm) buffer, Mo(1.0 nm) buffer and no buffer. Non-magnetic neutron and X-ray components are shown in (a) with magnetic polarized neutron spin asymmetry information shown in (b) as a function of an applied, in-plane field. Both are fitted by non-magnetic and magnetic scattering length density (SLD) distributions of which 1$\sigma$ confidence intervals are shown in (c) and (d) respectively.}
    \label{fig:PNR_all}
\end{figure}

\begin{figure}
    \centering
    \includegraphics[width=0.6\linewidth]{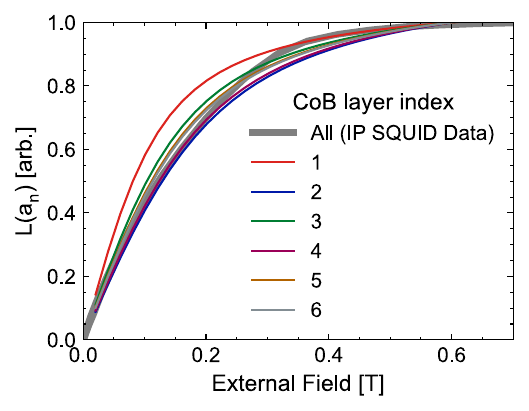}
    \caption{Modelled CoB layers' individual Langevin dependences in the \BS/Ta(2.0 nm)/MML structure compared to the stack-averaged SQUID magnetometry.}
    \label{fig:Langevins}
\end{figure}
\subsection{Slab Modelling}

The data was simulated using the Refl1D Python library and fitted using the DREAM algorithm. 2 reflectivity measurements at each of 5 (3 for non-buffered and Ta(15 \AA) buffer samples) fields alongside sample specific XRR curves were co-fitted. All models share the same structural parameters. The final model used modelled individual CoB layers with variable magnetisations idealised as 6 layer-by-layer Langevin functions. Note whilst these layers are magnetically independent they share thicknesses, densities and roughnesses. The combined X-ray, nuclear neutron and magnetic neutron SLDs are shown in figure \ref{fig:PNR_all} and the modelled IP Langevin relations compared to SQUID magnetometry are shown in figure \ref{fig:Langevins}.
  \begin{figure}
    \centering
    \includegraphics[width=\linewidth]{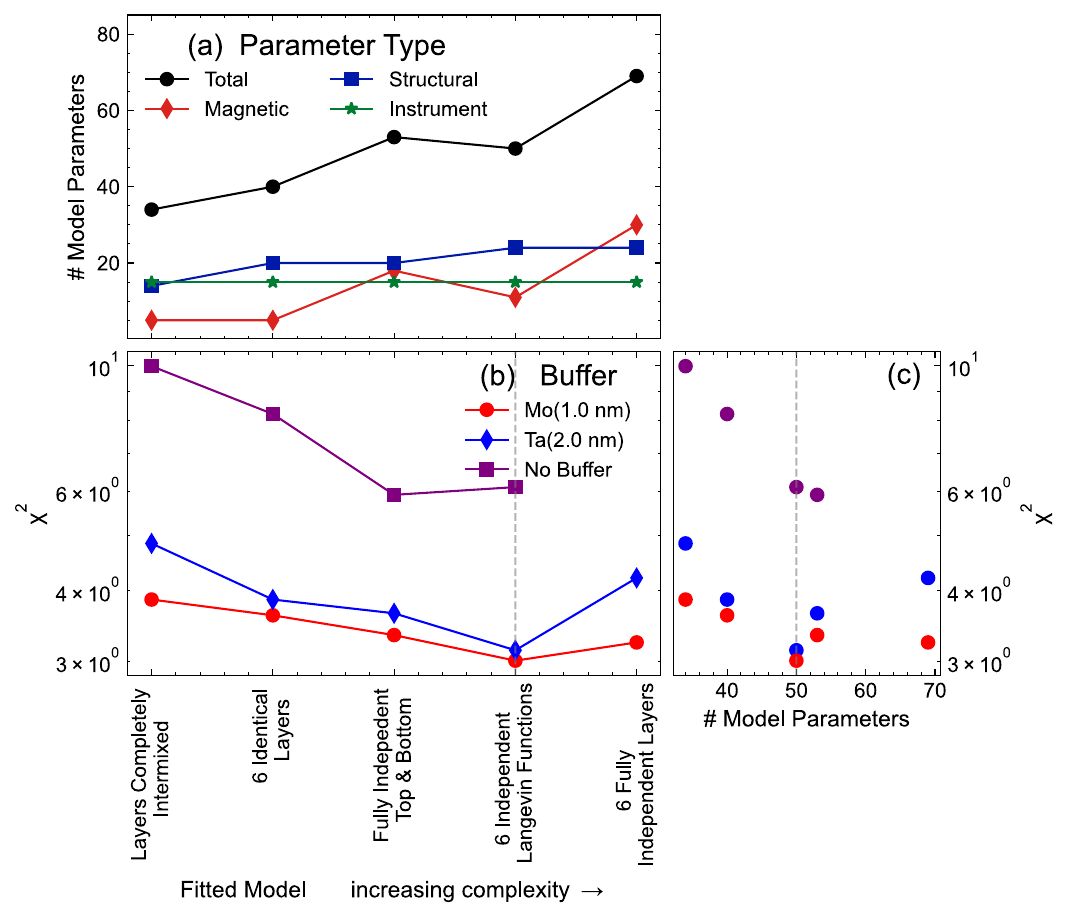}
    \caption{Comparison of the models used to fit neutron reflectivity data increasing in the complexity / freedom in magnetic modelling from left to right showing: (a) the number of parameters for each model by category; (b) The neutron $\chi^2$ dependence on selected model and (c) the neutron $\chi^2$ dependence as a function of increasing number of parameters. The dashed grey line denotes the final Langevin model used throughout and shown in Fig. \ref{fig:PNR_all} for the different buffers}
    \label{fig:chi2}
\end{figure}
 Shown in figure \ref{fig:chi2} are the chi-squared values for models of increasing complexity whilst fitting neutron data only. Whilst modelling the full multilayer as a single average magnetic layer obtains a satisfactory fit of $\chi^2 \sim 4$ for buffered samples,  modelling these layers as individually resolvable results in an improvement in the final reduced chi-squared test statistic indicating we are weakly sensitive to this lengthscale. A phenomenological `oxidation' parameter was introduced: modelled by decreasing the $M_s$ of the topmost layer by a fraction, This was found to be necessary to include otherwise the model would compensate by increasing $H_k$ such that the layer would not saturate at 0.7 T and is necessary for fitting the features in spin asymmetry at $Q \sim 0.065 $\AA$^{-1}$. It is notable however when this oxidation parameter is large ($\sim 50 \%$), the error in the extracted effective anisotropy of the topmost layer is greatly increased. 

 Structurally this can be improved further with the introduction of X-ray co-fitting. The X-rays' systematic errors, due to their much higher flux, tend to dominate the $\chi^2$ dependence and hence the fitting, for which an additional 10 \% error is included. Here the fully intermixed model only achieves $\chi^2 \sim 16$ whilst individual layers: $\chi^2 = 2.8$ for the Ta(2.0 nm) buffer. It should be noted the quality of the X-ray fit was improved dramatically when layer densities were allowed to take different values for neutrons and X-ray SLDs. Considering corresponding values usually lay within $\pm 5\%$ of each other this was not regarded as significant.   
 
 The fully intermixed model obtains $\chi^2 \sim 10$ where the heterostructure is grown without a buffer layer resulting in behaviour being anisotropic throughout the stack which improves to $\chi^2 = 6.1$ when modelling the interfacial layer perseverate. The weaker chi-squared dependence for buffered samples can hence be taken as a consequence of their uniform structure. 
 
 Further increasing the freedom and complexity in the fit model such that the magnetisation of each layer at each field is a completely free parameter does not improve the chi-squared suggesting too many parameters have been introduced for the model fitting to properly converge. This hence justifies coupling the layers magnetisation at each field under an effective magnetic `stiffness' parameter, $a_n$ related to $K_{\text{eff},n}$ and as the five magnetisations of any one layer at 5 fields are not independent of each other. Assuming the layers have full PMA we can approximate the effective anisotropy as the integral of the difference between modelled in-plane and measured out-of-plane curves:

 \begin{equation} 
     K_{\text{eff},n} \approx \mu_0\int^{H_{K}}_{H = 0}\left( M_{\text{SQUID, OOP}}-  M_s{\cal L}(a_nH) \right)\ dH  
 \end{equation}

\begin{figure}
    \centering
    \includegraphics[width=0.9\linewidth]{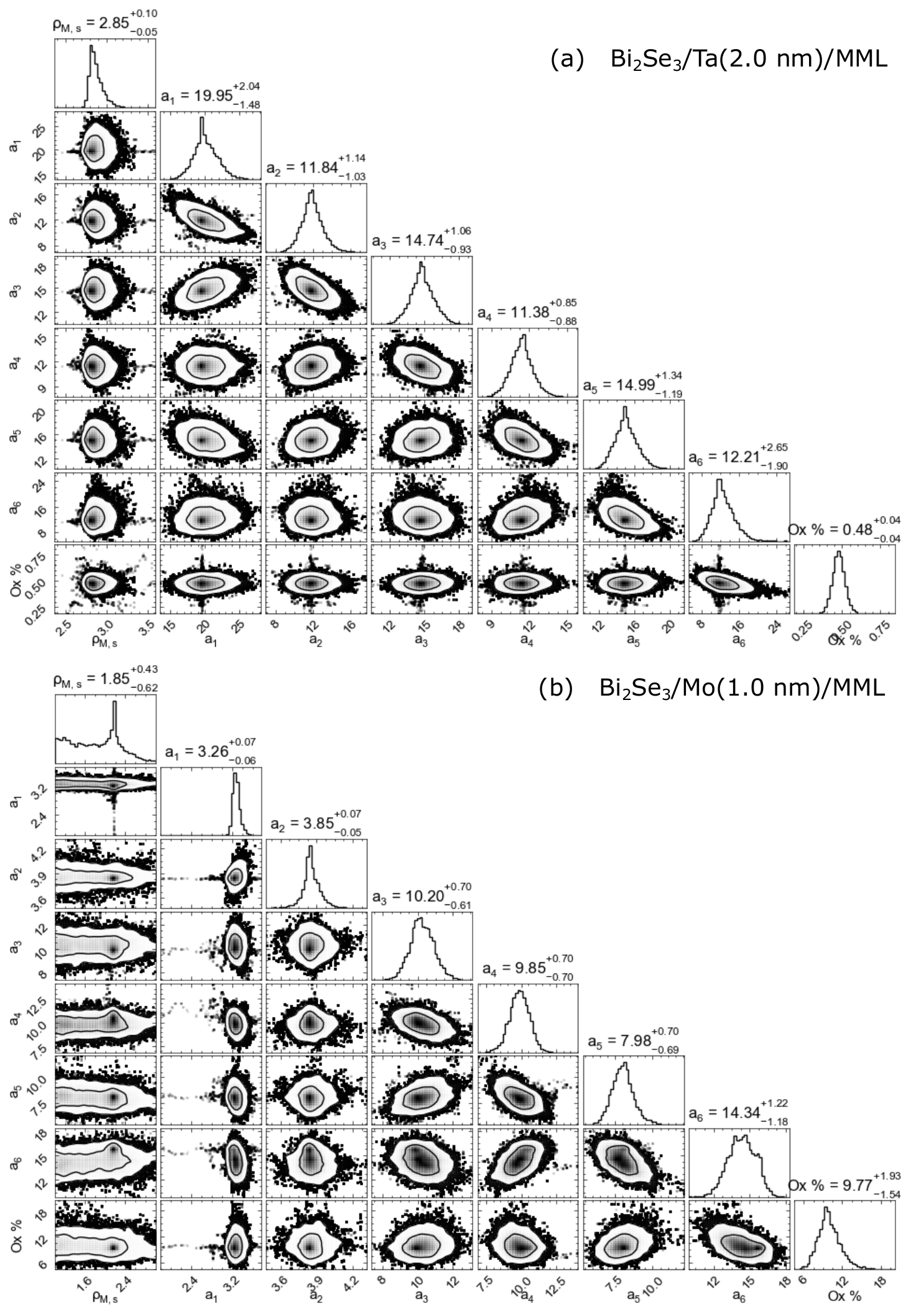}
    \caption{Bumps fit magnetic parameter correlations for Ta(2.0 nm) (a) and Mo(1.0) (b) buffers. Contour plots show both 68\% and 95\% confidence intervals. Plots created using the corner python package \cite{corner}.}
    \label{fig:correlations}
\end{figure}

\begin{figure}
    \centering
    \includegraphics[width=0.78\linewidth]{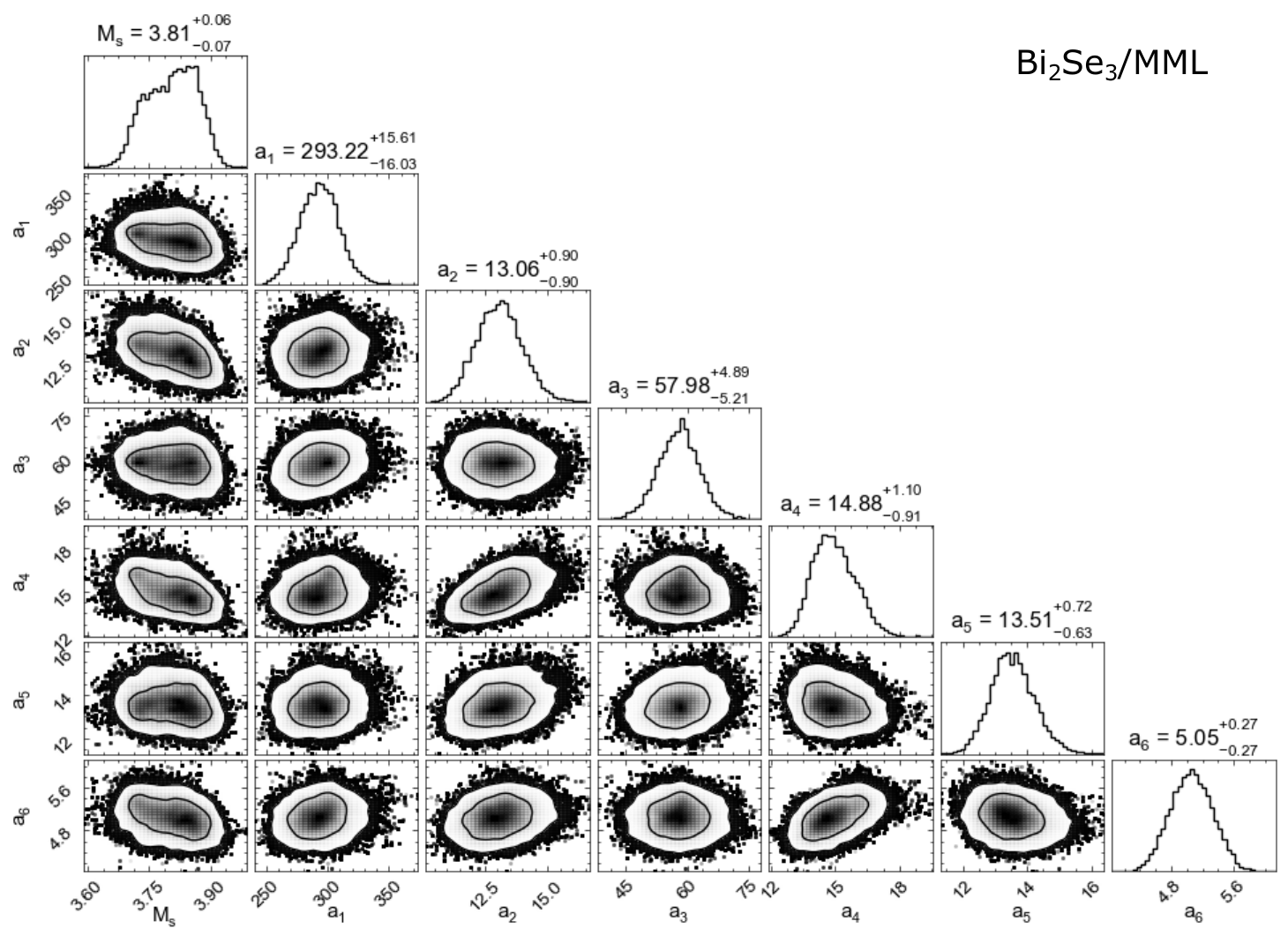}
    \caption{Bumps fit magnetic parameter correlations for sample without a buffer layer. Contour plots show both 68\% and 95\% confidence intervals. This sample has a 4 nm Pt cap and hence lacks an oxidation parameter. Plots created using the corner python package \cite{corner}.}
    \label{fig:correlations_nb}
\end{figure}

Looking at the magnetic parameter correlations in Figs.~\ref{fig:correlations} \& \ref{fig:correlations_nb} informs us on whether parameters are meaningful and independent. The majority of parameters appear uncorrelated, particularly for the non-buffered sample in Fig.~\ref{fig:correlations_nb}. For the two buffered samples in Fig.~\ref{fig:correlations} we notice weak to moderate correlation between the modelled $a_n$ values for neighbouring FM layers indicating individual layer modelling is partially limited due to our statistics at high $Q$. 

Neutron beams are known to typically possess a coherence length in the micron-scale, far exceeding the average \BS\ terrace width for both wide and narrow \BS\ terrace types. Hence we are able to assume the effective medium approximation applies and that in reality the presence of terracing from the \BS\ will blur the multilayer features over the lengthscale of $\sim 1$ nm. This likely renders individual layers to be not fully resolvable regardless of model choice or statistics collected.

All models used can be found in the open access repository \cite{data}.

\section{Multilayer Simulations}
\subsection{Micromagnetic Parameters}

All but the DMI vector strength are extracted / calculated from SQUID values. The DMI strength was chosen to be a value such that labyrinthine domains would form. The values used are shown in table \ref{tab:vals}. 

\begin{table}[]
    \centering
    \begin{tabular}{c|c|c|c}
       M$_S$ [MA/m]  &  A$_{\text{ex}}$ [J/m$^3$]  & K$_U$ [MJ/m$^3$]\\\hline  0.75 & 1.8 $\times 10^{-11}$ 
         & 1.00    \end{tabular}
    \caption{Micromagnetic Parameters exctracted from magnetometry used in model.}
    \label{tab:vals}
\end{table}

\begin{figure}
    \centering
    \includegraphics[width=0.6\linewidth]{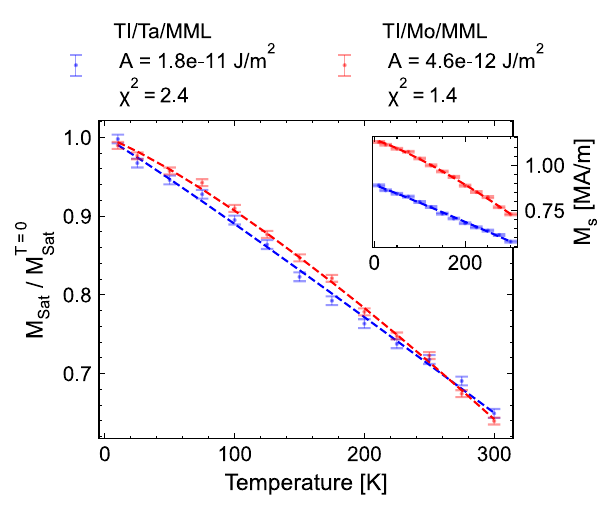}
    \caption{Fitted Thin film M v H models to SQUID-VSM data shown for two multilayer structures grown on wide-terraced Bi$_2$Se$_3$.}
    \label{fig:BlochLaw}
\end{figure}

Exchange stiffness was calculated from fitting a magnon dispersion relation in thin films. We here use a form based on the equations outlined in the supplementary section of Huang \textit{et al.} (2025) \cite{Huang2025}:

\begin{equation}
    m_z(T) =  1 + \frac{1}{k_BT}\frac{8\pi}{L_z}\frac{2g\mu_B}{M_SD}\sum^{N-1}_{m = 0} \ln \left[ 1 - \exp \left( \frac{g\mu_BB_0}{k_BT} - \frac{D}{k_BT}\left(\frac{m\pi}{(N-1)a}\right)^2\right) \right]
\end{equation}
where $L_z$, $D$, $B_0$ and $a$ are the fitted values corresponding to film width, magnon stiffness, external field and lattice parameter respectively. We use the conversion $A_{\text{ex}} = \frac{DM_S}{2g\mu_B}$ to obtain the exchange stiffness. M v T relations are fitted for both multilayers on wide-terraced Bi$_2$Se$_3$ with Ta(2.0 nm) and Mo(1.0 nm) buffers and are shown in Fig. \ref{fig:BlochLaw}. Using Bumps fitting,  a factor of $\sim$ 4 difference is found between them. Both fits achieve satisfactory chi-squared values. For simulations, the values extracted from the sample on the Ta buffer were chosen. Mumax3 uses uniaxial magnetic anisotropy, $K_U$ where $K_U = K_{\text{eff}} + \frac{1}{2}\mu_0M_S^2 = K_{\text{eff}} + 0.34 ~\text{MJ/m}^3$

\subsection{Micromagnetic Model}

Simulations composed of 6, 0.6 nm magnetic layers separated by non magnetic spaces 1.2 nm layers thick and were simulated using mumax3 \cite{Vansteenkiste2014, Mulkers2017}. Discretisation in $x,y$ is set to 2.5 nm to be below the key lengthscales of the wall-width, $\delta = \sqrt{\frac{A}{K}} \sim 5$ nm and exchange length, $l_{\text{ex}} =\sqrt{\frac{A} {\mu_0M_s^2}} \sim 5$ nm. The model used can be found in the open access repository \cite{data}. 

The DMI vector strength $D$ was increased such that a labyrinthine maze domain state formed reliably through lowering domain wall energy. This was found to be D = 5 meV/m$^2$. This is likely multiple times higher than the true value and the domain nucleation energy in the true films will instead be lowered by the presence of grains and disorder however this is beyond the scope of this paper. The simulated domains are an order of magnitude than those observed ($\sim 10 - 30$ nm instead of $\sim 400 - 600$ nm) however are included as a test that arbitrary spin textures couple magneto-statically which they are observed to do so.  
\begin{figure}
    \centering
    \includegraphics[width=0.9\linewidth]{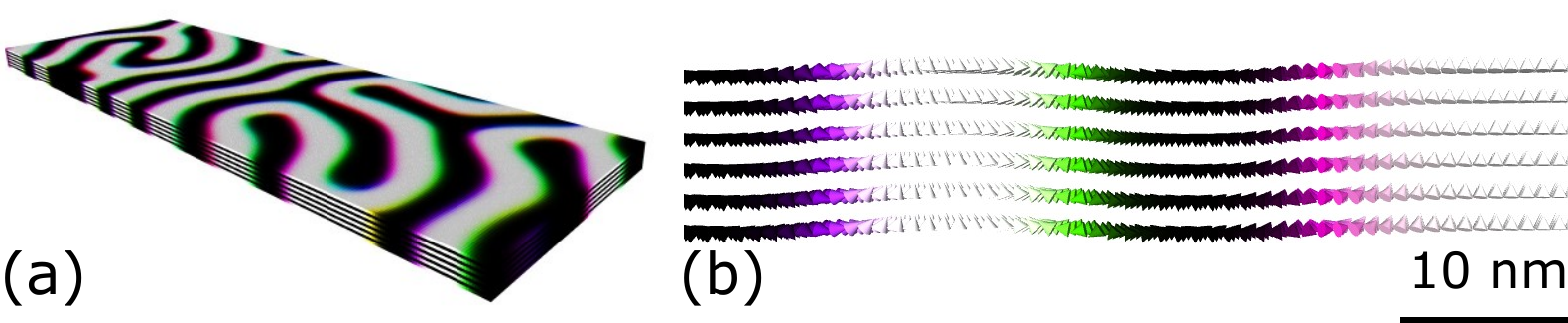}
    \caption{(a) DMI-induced spin textures coupling magnetostatically throughout the multilayer; (b) cross sectional view.}
    \label{fig:MMs}
\end{figure}
\bibliographystyle{abbrv}
\bibliography{library}